\documentclass[thmsa]{article}
\usepackage{sw20lart}

\input tcilatex
\QQQ{Language}{
British English
}

\begin{document}

\section{Introduction}

In a recent paper $[1]$, we have shown how the interpretation of quantum
mechanics due to Land\'e $[2-5]$ can be used to derive spin operators and
their eigenvectors from first principles. This interpretation allows us to
go further than has to date been done in the treatment of spin and
ultimately enables us to obtain new generalized forms for the spin operators
and their eigenvectors. In $[1]$, we applied this treatment to the case of
spin $1/2$ and derived the generalized formulas for $\left[ \sigma _x\right] 
$, $[\sigma _y]$ and $\left[ \sigma _z\right] $, the $x$, $y$ and $z$
components of spin, and for the eigenvectors of $\left[ \sigma _z\right] .$
The operators $\left[ \sigma _x\right] $ and $[\sigma _y]$ were obtained
through the ladder operators, which in turn were derived from their actions
on the eigenvectors of $[\sigma _z].$ However, the generalized eigenvectors
of $\left[ \sigma _x\right] $ and $[\sigma _y]$ were not given. In this
paper, we present these eigenvectors. Furthermore, we introduce a method for
obtaining $\left[ \sigma _x\right] $ and $[\sigma _y]$ and their
eigenvectors without the agency of the ladder operators. This is very
important because the method involving the ladder operators is quite
tedious, as evidenced even for the simplest case spin $1/2$. Finally, we
prove that the generalized operator for the square of the spin is a unit
matrix multiplied by the value of the square of the spin.

\section{Basic Theory}

The method we have devised and used to obtain the spin-$1/2$ quantities is
founded on the Land\'e interpretation of quantum mechanics $[2-5]$. Let $A$, 
$B$ and $C$ be observables of a quantum system. The eigenvalues of $A$ are $%
A_1$, $A_2$ ... The eigenvalues of $B$ are $B_1$, $B_2$ .... Finally, the
eigenvalues of $C$ are $C_1$, $C_2$, ... Suppose that the initial state of
the system corresponds to $A_i.$ Starting from this state, let the
probability amplitude for obtaining an eigenvalue $C_n$ upon measurement of $%
C$ be $\psi \left( A_i;C_n\right) .$ Let the probability amplitude for
obtaining the eigenvalue $B_j$ of $B$ be $\chi (A_i;B_j)$. Finally let the
probability amplitude for obtaining the eigenvalue $C_n$ of $C$ when the
system is initially in a state corresponding to $B_j$ be $\phi (B_j;C_n).$
Then the three sets of probability amplitudes are connected by the formula $%
[2,3]$

\begin{equation}
\psi \left( A_i;C_n\right) =\sum_j\chi (A_i;B_j)\phi (B_j;C_n).
\label{fo40a}
\end{equation}
These probability amplitudes obey two-way symmetry contained in the
Hermiticity condition

\begin{equation}
\psi \left( A_i;C_n\right) =\psi ^{*}\left( C_n;A_i\right) ,\text{etc.}
\label{fo41a}
\end{equation}
They also satisfy

\begin{equation}
\zeta (A_i;A_j)=\delta _{ij},\;\zeta (B_i;B_j)=\delta _{ij},\text{ etc,}
\label{fo42a}
\end{equation}
which simply ensures the repeatability of measurement.

For spin, the three observables $A$, $B$ and $C$ correspond to projection
measurements along three different quantization axes. This means that $\psi
, $ $\chi $ and $\phi $ are identical in form.

Suppose the quantity $R$ is a function of the final spin projection
resulting from spin measurement. In matrix mechanics, $R$ is represented by
the spin operator $[R]$. In $[1]$, we derived the general expressions for
the matrix elements of $[R]$ for the case of spin $1/2$. Thus, for this case
spin $1/2$ let the initial spin projection be with respect to the unit
vector $\widehat{{\bf a}}$ and let the final spin projection be with respect
to the unit vector $\widehat{{\bf c}}$. Let the polar angles of $\widehat{%
{\bf a}}$ be $(\theta ^{\prime \prime },\varphi ^{\prime \prime })$ and the
polar angles of $\widehat{{\bf c}}$ be $(\theta ^{\prime },\varphi ^{\prime
})$. Let the probability amplitudes for measurements of spin from $\widehat{%
{\bf a}}$ to $\widehat{{\bf c}}$ be $\psi (m_i^{(\widehat{{\bf a}})};m_f^{(%
\widehat{{\bf c}})})$, with $m_i,m_f=1,2$ and $m_1^{(\widehat{{\bf a}}%
)}=m_1^{(\widehat{{\bf c}})}=+\frac 12$, while $m_2^{(\widehat{{\bf a}}%
)}=m_2^{(\widehat{{\bf c}})}=-\frac 12.$ For example, $\psi ((+\frac 12)^{(%
\widehat{{\bf a}})};(-\frac 12)^{(\widehat{{\bf c}})})$ is the probability
amplitude for finding the spin projection to be down with respect to $%
\widehat{{\bf c}}$ when the system was initially in a state in which the
spin projection was up with respect to $\widehat{{\bf a}}$.

Let the quantity $R$ take the values $r_1$ and $r_2$ when the spin
projection is found up and down respectively with respect to the final
direction. Then the elements of the matrix representatiion 
\begin{equation}
\lbrack R]=\left( 
\begin{array}{cc}
R_{11} & R_{12} \\ 
R_{21} & R_{22}
\end{array}
\right)  \label{th31}
\end{equation}
of $R$ are $[1]$ 
\begin{equation}
R_{11}=\left| \phi ((+\frac 12)^{(\widehat{{\bf b}})};(+\frac 12)^{(\widehat{%
{\bf c}})})\right| ^2r_1+\left| \phi ((+\frac 12)^{(\widehat{{\bf b}}%
)};(-\frac 12)^{(\widehat{{\bf c}})})\right| ^2r_2  \label{th32}
\end{equation}

\begin{eqnarray}
R_{12} &=&\phi ^{*}((+\frac 12)^{(\widehat{{\bf b}})};(+\frac 12)^{(\widehat{%
{\bf c}})})\phi ((-\frac 12)^{(\widehat{{\bf b}})};(+\frac 12)^{(\widehat{%
{\bf c}})})r_1  \nonumber \\
&&\ +\phi ^{*}((+\frac 12)^{(\widehat{{\bf b}})};(-\frac 12)^{(\widehat{{\bf %
c}})})\phi ((-\frac 12)^{(\widehat{{\bf b}})};(-\frac 12)^{(\widehat{{\bf c}}%
)})r_2,  \label{th33}
\end{eqnarray}

\begin{eqnarray}
R_{21} &=&\phi ^{*}((-\frac 12)^{(\widehat{{\bf b}})};(+\frac 12)^{(\widehat{%
{\bf c}})})\phi ((+\frac 12)^{(\widehat{{\bf b}})};(+\frac 12)^{(\widehat{%
{\bf c}})})r_1  \nonumber \\
&&\ +\phi ^{*}((-\frac 12)^{(\widehat{{\bf b}})};(-\frac 12)^{(\widehat{{\bf %
c}})})\phi ((+\frac 12)^{(\widehat{{\bf b}})};(-\frac 12)^{(\widehat{{\bf c}}%
)})r_2  \label{th34}
\end{eqnarray}
and

\begin{equation}
R_{22}=\left| \phi ((-\frac 12)^{(\widehat{{\bf b}})};(+\frac 12)^{(\widehat{%
{\bf c}})})\right| ^2r_1+\left| \phi ((-\frac 12)^{(\widehat{{\bf b}}%
)};(-\frac 12)^{(\widehat{{\bf c}})})\right| ^2r_2  \label{th35}
\end{equation}

Here $\phi (m_i^{(\widehat{{\bf b}})};m_j^{(\widehat{{\bf c}})})$ are the
probability amplitudes for the outcomes of spin projection measurements from
the direction $\widehat{{\bf b}}$ to the direction $\widehat{{\bf c}},$
where $\widehat{{\bf b}}$ is an intermediate quantization direction defined
by the angles $(\theta ,\varphi ).$

\section{Spin Operators and Eigenvectors}

Suppose that the quantity $R$ is the spin projection itself, which we
measure in units of $\hbar /2$. If the initial spin state is characterized
by the magnetic quantum number $m_i^{(\widehat{{\bf a}})}$ defined with
respect to the direction $\widehat{{\bf a}}$, then the expectation value of
the spin projection along the direction $\widehat{{\bf c}}$ is $[1]$

\begin{equation}
\left\langle \sigma _{\widehat{{\bf c}}}\right\rangle =[\psi (m_i^{(\widehat{%
{\bf a}})})]^{\dagger }[\sigma _{\widehat{{\bf c}}}][\psi (m_i^{(\widehat{%
{\bf a}})})],  \label{onea}
\end{equation}
where

\begin{equation}
\lbrack \sigma _{\widehat{{\bf c}}}]=\left( 
\begin{array}{cc}
(\sigma _{\widehat{{\bf c}}})_{11} & (\sigma _{\widehat{{\bf c}}})_{12} \\ 
(\sigma _{\widehat{{\bf c}}})_{21} & (\sigma _{\widehat{{\bf c}}})_{22}
\end{array}
\right) ,  \label{one}
\end{equation}
and 
\begin{equation}
\lbrack \psi (m_i^{(\widehat{{\bf a}})})]=\left( 
\begin{array}{c}
\chi (m_i^{(\widehat{{\bf a}})};(+\frac 12)^{(\widehat{{\bf b}})}) \\ 
\chi (m_i^{(\widehat{{\bf a}})};(-\frac 12)^{(\widehat{{\bf b}})})
\end{array}
\right)   \label{onez}
\end{equation}
is the vector state corresponding to the initial spin projection $m_i^{(%
\widehat{{\bf a}})}\hbar $ along $\widehat{{\bf a}}.$

The elements of $[\sigma _{\widehat{{\bf c}}}]$ are functions of the angles $%
(\theta ,\varphi )$ of $\widehat{{\bf b}}$ and the angles $(\theta ^{\prime
},\varphi ^{\prime })$ of $\widehat{{\bf c}}$. They are $[1]$%
\begin{equation}
(\sigma _{\widehat{{\bf c}}})_{11}=\cos \theta \cos \theta ^{\prime }+\sin
\theta \sin \theta ^{\prime }\cos (\varphi -\varphi ^{\prime }),  \label{two}
\end{equation}
\begin{equation}
(\sigma _{\widehat{{\bf c}}})_{12}=\sin \theta \cos \theta ^{\prime }-\sin
\theta ^{\prime }[\cos \theta \cos (\varphi -\varphi ^{\prime })+i\sin
(\varphi -\varphi ^{\prime })],  \label{three}
\end{equation}
\begin{equation}
(\sigma _{\widehat{{\bf c}}})_{21}=\sin \theta \cos \theta ^{\prime }-\sin
\theta ^{\prime }[\cos \theta \cos (\varphi -\varphi ^{\prime })-i\sin
(\varphi -\varphi ^{\prime })]  \label{four}
\end{equation}
and 
\begin{equation}
(\sigma _{\widehat{{\bf c}}})_{22}=-\cos \theta \cos \theta ^{\prime }-\sin
\theta \sin \theta ^{\prime }\cos (\varphi -\varphi ^{\prime }).
\label{five}
\end{equation}

This operator $[\sigma _{\widehat{{\bf c}}}]\;$is the component of the spin
along the direction $\widehat{{\bf c}}$. This is evidently a more
generalized form of this operator than

\begin{equation}
\lbrack {\bf \sigma }\cdot \widehat{{\bf c}}]=\left( 
\begin{array}{cc}
\cos \theta ^{\prime } & \sin \theta ^{\prime }e^{-i\varphi ^{\prime }} \\ 
\sin \theta ^{\prime }e^{i\varphi ^{\prime }} & -\cos \theta ^{\prime }
\end{array}
\right) ,  \label{fivey}
\end{equation}
found in the literature. Here 
\begin{equation}
\lbrack {\bf \sigma }]=\widehat{{\bf i}}[\sigma _x]+\widehat{{\bf j}}[\sigma
_y]+\widehat{{\bf k}}[\sigma _z]  \label{fivez}
\end{equation}
where $[\sigma _x],\;[\sigma _y]\;$and $[\sigma _z]$ are the Pauli spin
matrices.

In the literature Eqn. (\ref{fivez}) is presented as being the generalized
form of the $z$ component of the spin, since it refers to the arbitrary
direction $\widehat{{\bf c}}$. But this form can be obtained from the
generalized form Eqs. (\ref{two}) - (\ref{five}) by setting $\theta =\varphi
=0;$ this corresponds to the vector $\widehat{{\bf b}}$ pointing in the
positive $z$ direction. In the limit $\theta =\theta ^{\prime }$ and $%
\varphi =\varphi ^{\prime }$, we obtain from the generalized form the Pauli
matrix

\begin{equation}
\lbrack \sigma _z]=\left( 
\begin{array}{cc}
1 & 0 \\ 
0 & -1
\end{array}
\right) .  \label{fived}
\end{equation}
Moreover, the generalized $x$ and $y$ components of spin which we obtained
in $[1]$, and which we shall derive in an alternative way below, change to
the corresponding Pauli matrices in the same limit.

The spin state $[\psi _{+}]$ is $[1]$

\begin{equation}
\lbrack \psi ((+\tfrac 12)^{(\widehat{{\bf a}})})]=\left( 
\begin{array}{c}
\cos \theta ^{\prime \prime }/2\cos \theta /2+e^{i(\varphi ^{\prime \prime
}-\varphi )}\sin \theta ^{\prime \prime }/2\sin \theta /2 \\ 
\cos \theta ^{\prime \prime }/2\sin \theta /2-e^{i(\varphi ^{\prime \prime
}-\varphi )}\sin \theta ^{\prime \prime }/2\cos \theta /2
\end{array}
\right) .  \label{fivea}
\end{equation}
This state is evidently the most generalized form of the spin state.

If on the other hand the spin projection is initially down with respect to $%
\widehat{{\bf a}}$, the vector $[\psi _{+}]$ is replaced by $[1]$ 
\begin{equation}
\lbrack \psi ((-\tfrac 12)^{(\widehat{{\bf a}})})]=\left( 
\begin{array}{c}
\sin \theta ^{\prime \prime }/2\cos \theta /2-e^{i(\varphi ^{\prime \prime
}-\varphi )}\cos \theta ^{\prime \prime }/2\sin \theta /2 \\ 
\sin \theta ^{\prime \prime }/2\sin \theta /2+e^{i(\varphi ^{\prime \prime
}-\varphi )}\cos \theta ^{\prime \prime }/2\cos \theta /2
\end{array}
\right) .  \label{fiveb}
\end{equation}
in the expression Eq. (\ref{onea}) for the expectation value.

The eigenvalues of $[\sigma _{\widehat{{\bf c}}}]$ are $+1$ with the
eigenvector $[1$]

\begin{equation}
\lbrack \xi _{\widehat{{\bf c}}}^{(+)}]=\left( 
\begin{array}{c}
\phi ((+\frac 12)^{(\widehat{{\bf c}})};(+\frac 12)^{(\widehat{{\bf b}})})
\\ 
\phi ((+\frac 12)^{(\widehat{{\bf c}})};(-\frac 12)^{(\widehat{{\bf b}})})
\end{array}
\right) =\left( 
\begin{array}{c}
\cos \theta ^{\prime }/2\cos \theta /2+e^{i(\varphi ^{\prime }-\varphi
)}\sin \theta ^{\prime }/2\sin \theta /2 \\ 
\cos \theta ^{\prime }/2\sin \theta /2-e^{i(\varphi ^{\prime }-\varphi
)}\sin \theta ^{\prime }/2\cos \theta /2
\end{array}
\right)  \label{six}
\end{equation}
and $-1$ with the eigenvector [1]

\begin{equation}
\lbrack \xi _{\widehat{{\bf c}}}^{(-)}]=\left( 
\begin{array}{c}
\phi ((-\frac 12)^{(\widehat{{\bf c}})};(+\frac 12)^{(\widehat{{\bf b}})})
\\ 
\phi ((-\frac 12)^{(\widehat{{\bf c}})};(-\frac 12)^{(\widehat{{\bf b}})})
\end{array}
\right) =\left( 
\begin{array}{c}
\sin \theta ^{\prime }/2\cos \theta /2-e^{i(\varphi ^{\prime }-\varphi
)}\cos \theta ^{\prime }/2\sin \theta /2 \\ 
\sin \theta ^{\prime }/2\sin \theta /2+e^{i(\varphi ^{\prime }-\varphi
)}\cos \theta ^{\prime }/2\cos \theta /2
\end{array}
\right) .  \label{seven}
\end{equation}

We also derived the generalized operators $\left[ \sigma _x\right] $ and $%
[\sigma _y].$ These operators correspond to directions such that in the
limit $\theta =\theta ^{\prime }$ and $\varphi =\varphi ^{\prime }$, they
become the Pauli matrices $\left[ \sigma _x\right] $ and $[\sigma _y].$ The
elements of $\left[ \sigma _x\right] $ are $[1]$ 
\begin{equation}
\left( \sigma _x\right) _{11}=-\sin \theta \cos \theta ^{\prime }\cos
(\varphi ^{\prime }-\varphi )+\sin \theta ^{\prime }\cos \theta ,
\label{eight}
\end{equation}
\begin{equation}
(\sigma _x)_{12}=\cos \theta \cos \theta ^{\prime }\cos (\varphi ^{\prime
}-\varphi )+\sin \theta \sin \theta ^{\prime }-i\cos \theta ^{\prime }\sin
(\varphi ^{\prime }-\varphi ),  \label{nine}
\end{equation}

\begin{equation}
(\sigma _x)_{21}=\cos \theta \cos \theta ^{\prime }\cos (\varphi ^{\prime
}-\varphi )+\sin \theta \sin \theta ^{\prime }+i\cos \theta ^{\prime }\sin
(\varphi ^{\prime }-\varphi ),  \label{ten}
\end{equation}
and 
\begin{equation}
\left( \sigma _x\right) _{22}=\sin \theta \cos \theta ^{\prime }\cos
(\varphi ^{\prime }-\varphi )-\sin \theta ^{\prime }\cos \theta ,
\label{el11}
\end{equation}
while the elements of $[\sigma _y]$ are 
\begin{equation}
(\sigma _y)_{11}=\sin \theta \sin (\varphi ^{\prime }-\varphi ),
\label{tw12}
\end{equation}
\begin{equation}
(\sigma _y)_{12}=-\cos \theta \sin (\varphi ^{\prime }-\varphi )-i\cos
(\varphi ^{\prime }-\varphi ),  \label{th13}
\end{equation}
\begin{equation}
(\sigma _y)_{21}=-\cos \theta \sin (\varphi ^{\prime }-\varphi )+i\cos
(\varphi ^{\prime }-\varphi )  \label{fo14}
\end{equation}
and

\begin{equation}
(\sigma _y)_{22}=-\sin \theta \sin (\varphi ^{\prime }-\varphi ).
\label{fi15}
\end{equation}

The eigenvectors of $\left[ \sigma _x\right] $ and $[\sigma _y]$ were not
given.

\section{Alternative Derivation of the x and y Components of the Spin}

We have found that we can obtain the $x$ and $y$ components of the spin
operator by changing the arguments appropriately in the formula for the $z$
component. This same change of argument applied to the eigenvectors of the $%
z $ component yields the eigenvectors of the $x$ and $y$ components.

The change of argument $\theta ^{\prime }\rightarrow \theta ^{\prime }-\pi
/2 $ in the formulas Eqs. \ref{two} - \ref{five} leads to elements of $%
[\sigma _x].$ Thus

\begin{equation}
(\sigma _{\widehat{{\bf c}}})_{11}\rightarrow \cos \theta \sin \theta
^{\prime }-\sin \theta \cos \theta ^{\prime }\cos (\varphi -\varphi ^{\prime
})=(\sigma _x)_{11}  \label{si16}
\end{equation}
\begin{equation}
(\sigma _{\widehat{{\bf c}}})_{12}\rightarrow \cos \theta \cos \theta
^{\prime }\cos (\varphi ^{\prime }-\varphi )+\sin \theta \sin \theta
^{\prime }-i\cos \theta ^{\prime }\sin (\varphi ^{\prime }-\varphi )=(\sigma
_x)_{12},  \label{se17}
\end{equation}

\begin{equation}
(\sigma _{\widehat{{\bf c}}})_{21}\rightarrow \cos \theta \cos \theta
^{\prime }\cos (\varphi ^{\prime }-\varphi )+\sin \theta \sin \theta
^{\prime }+i\cos \theta ^{\prime }\sin (\varphi ^{\prime }-\varphi )=(\sigma
_x)_{21}  \label{ei18}
\end{equation}
and 
\begin{equation}
(\sigma _{\widehat{{\bf c}}})_{22}\rightarrow \sin \theta \cos \theta
^{\prime }\cos (\varphi ^{\prime }-\varphi )-\sin \theta ^{\prime }\cos
\theta =\left( \sigma _x\right) _{22}.  \label{ni19}
\end{equation}

These are just the same expressions as Eq. (\ref{eight}) - (\ref{el11}). The
same transformation should yield the eigenvectors of $[\sigma _x]$ if
applied to those for $[\sigma _{\widehat{{\bf c}}}].$ Under this change of
argument, we obtain

\begin{equation}
\lbrack \xi _{\widehat{{\bf c}}}^{(+)}]\rightarrow [\xi _x^{(+)}]=\left( 
\begin{array}{c}
\frac 1{\sqrt{2}}[(\sin \theta ^{\prime }/2+\cos \theta ^{\prime }/2)\cos
\theta /2+e^{i(\varphi ^{\prime }-\varphi )}(\sin \theta ^{\prime }/2-\cos
\theta ^{\prime }/2)\sin \theta /2] \\ 
\frac 1{\sqrt{2}}[(\sin \theta ^{\prime }/2+\cos \theta ^{\prime }/2)\sin
\theta /2-e^{i(\varphi ^{\prime }-\varphi )}(\sin \theta ^{\prime }/2-\cos
\theta ^{\prime }/2)\cos \theta /2]
\end{array}
\right)
\end{equation}
and 
\begin{equation}
\lbrack \xi _{\widehat{{\bf c}}}^{(-)}]\rightarrow [\xi _x^{(-)}]=\left( 
\begin{array}{c}
\frac 1{\sqrt{2}}[(\sin \theta ^{\prime }/2-\cos \theta ^{\prime }/2)\cos
\theta /2-e^{i(\varphi ^{\prime }-\varphi )}(\sin \theta ^{\prime }/2+\cos
\theta ^{\prime }/2)\sin \theta /2] \\ 
\frac 1{\sqrt{2}}[(\sin \theta ^{\prime }/2-\cos \theta ^{\prime }/2)\sin
\theta /2+e^{i(\varphi ^{\prime }-\varphi )}(\sin \theta ^{\prime }/2+\cos
\theta ^{\prime }/2)\cos \theta /2]
\end{array}
\right) .  \label{tw21}
\end{equation}

Direct calculation shows that

\begin{equation}
\lbrack \sigma _x][\xi _x^{(\pm )}]=(\pm 1)[\xi _x^{(\pm )}].  \label{tw22}
\end{equation}

For the $y$ component, the transformations required are $\theta ^{\prime
}=\pi /2$ and $\varphi ^{\prime }\rightarrow \varphi ^{\prime }-\pi /2.$
Under these transformations, we obtain

\begin{equation}
(\sigma _{\widehat{{\bf c}}})_{11}\rightarrow \sin \theta \sin (\varphi
^{\prime }-\varphi )=(\sigma _y)_{11},  \label{tw24}
\end{equation}
\begin{equation}
(\sigma _{\widehat{{\bf c}}})_{12}\rightarrow \cos \theta \sin (\varphi
^{\prime }-\varphi )-i\cos (\varphi ^{\prime }-\varphi )=(\sigma _y)_{12},
\label{tw25}
\end{equation}
$,$%
\begin{equation}
(\sigma _{\widehat{{\bf c}}})_{21}\rightarrow \cos \theta \sin (\varphi
^{\prime }-\varphi )+i\cos (\varphi ^{\prime }-\varphi )=(\sigma _y)_{21}
\label{tw25a}
\end{equation}
and 
\begin{equation}
(\sigma _{\widehat{{\bf c}}})_{22}\rightarrow -\sin \theta \sin (\varphi
^{\prime }-\varphi )=\left( \sigma _x\right) _{22}.  \label{tw26}
\end{equation}

The eigenvectors of $[\sigma _{\widehat{{\bf c}}}]$ transform to

\begin{equation}
\lbrack \xi _{\widehat{{\bf c}}}^{(+)}]\rightarrow [\xi _y^{(+)}]=\left( 
\begin{array}{c}
\frac 1{\sqrt{2}}[\cos \theta /2-ie^{i(\varphi ^{\prime }-\varphi )}\sin
\theta /2] \\ 
\frac 1{\sqrt{2}}[\sin \theta /2+ie^{i(\varphi ^{\prime }-\varphi )}\cos
\theta /2]
\end{array}
\right)  \label{tw27}
\end{equation}
and 
\begin{equation}
\lbrack \xi _{\widehat{{\bf c}}}^{(-)}]\rightarrow [\xi _y^{(-)}]=\left( 
\begin{array}{c}
\frac 1{\sqrt{2}}[\cos \theta /2+ie^{i(\varphi ^{\prime }-\varphi )}\sin
\theta /2] \\ 
\frac 1{\sqrt{2}}[\sin \theta /2-ie^{i(\varphi ^{\prime }-\varphi )}\cos
\theta /2]
\end{array}
\right) .  \label{tw28}
\end{equation}

Again direct calculation proves that

\begin{equation}
\lbrack \sigma _y][\xi _y^{(\pm )}]=(\pm 1)[\xi _y^{(\pm )}].  \label{tw29}
\end{equation}

\section{Interpretation of the Components of the Spin}

The procedure for obtaining the $"x"$ and $"y"$ components of the spin
operator from the $"z"$ component may be summarized as follows. To obtain
the $x$ component, we change the final quantization direction in the
expression for the operator from $\widehat{{\bf c}}$ (whose polar angles are 
$(\theta ^{\prime },\varphi ^{\prime })$) to the direction $\widehat{{\bf c}}%
_x$ defined by the polar angles $(\theta ^{\prime }-\frac \pi 2,\varphi
^{\prime }).$ In order to obtain the $y$ component we have to change the
final quantization direction from $\widehat{{\bf c}}$ to the vector $%
\widehat{{\bf c}}_y$ whose angles are $(\theta ^{\prime }=\frac \pi
2,\varphi ^{\prime }-\frac \pi 2).$ The three vectors are

\begin{equation}
\widehat{{\bf c}}=(\sin \theta ^{\prime }\cos \varphi ^{\prime },\sin \theta
^{\prime }\sin \varphi ,\cos \theta ^{\prime })  \label{fo44}
\end{equation}

\begin{equation}
\widehat{{\bf c}}_x=(-\cos \theta ^{\prime }\cos \varphi ^{\prime },-\cos
\theta ^{\prime }\sin \varphi ^{\prime },\sin \theta ^{\prime })
\label{fo45}
\end{equation}
and

\[
\widehat{{\bf c}}_y=(\sin \varphi ^{\prime },-\cos \varphi ^{\prime },0). 
\]
They are mutually orthogonal:

\begin{equation}
\widehat{{\bf c}}\cdot \widehat{{\bf c}}_x=\widehat{{\bf c}}\cdot \widehat{%
{\bf c}}_y=\widehat{{\bf c}}_y\cdot \widehat{{\bf c}}=0.  \label{fo46}
\end{equation}
Moreover, their vector products are

\begin{equation}
\widehat{{\bf c}}_x\times \widehat{{\bf c}}_y=\widehat{{\bf c}};\;\;\;\;\;\;%
\widehat{{\bf c}}_y\times \widehat{{\bf c}}=\widehat{{\bf c}}_x;\;\;\;%
\widehat{{\bf c}}\times \widehat{{\bf c}}_x=\widehat{{\bf c}}_y\;\;\;\;
\label{fo47}
\end{equation}
Hence, these vectors form a new coordinate system whose $z$ axis is defined
by $\widehat{{\bf c}}$, the $x$ axis by $\widehat{{\bf c}}_x$ and the $y$
axis by $\widehat{{\bf c}}_y.$ These coordinate axes are rotated with
respect to the original ones. We may use this information to characterize
and interpret the components of spin.

Suppose the spin projection of a particle is known to be up or down along
the quantization direction $\widehat{{\bf a}}$. We then measure it along
direction $\widehat{{\bf c}}.$ If we seek the expectation value of the spin
projection for this situation, and convert the expression for this
expectation value to matrix form by means of the expansion Eq. (\ref{fo40a})
using the intermediate quantization direction $\widehat{{\bf b}}$, then the
matrix operator that results is the $z$ component of spin. The vector $%
\widehat{{\bf c}}$ then defines the $z$ axis of a new rotated coordinate
system.

If instead, starting from the same situation, we make a measurement of the
spin projection along the quantization direction defined by the positive $x$
axis of the new rotated coordinate system, then the operator that occurs in
the matrix expression for the expectation value is the $x$ component of
spin. The rotated $x$ axis is defined by the unit vector $\widehat{{\bf c}}%
_x $ whose angles are $(\theta ^{\prime }-\frac \pi 2,\varphi ^{\prime }).$
It lies in the same plane as the vector $\widehat{{\bf c}}$ and the original 
$z$ axis.

The rotated coordinate system has a $y$ axis defined by the angles $(\frac
\pi 2,\varphi ^{\prime }-\frac \pi 2).$ If we measure the spin projection
along this direction and compute the expectation value by the matrix
mechanics formula, using the intermediate quantization axis $\widehat{{\bf b}%
}$, the operator that features is the $y$ component of spin .

\section{Generalized Form for the Square of the Spin}

We obtain the elements of the matrix for the square of the spin by squaring
the components of the spin and adding them. We now proceed to show that this
labour is unnecessary, and that we can obtain these elements by a more
general method.

Since the spin is being measured in units of $\hbar /2,$, the eigenvalue of $%
\sigma ^2$ is $r_1=r_2=3.$ Hence, using Eqs. (\ref{th32}) - (\ref{th35}), we
obtain

\begin{equation}
(\sigma ^2)_{11}=3[\left| \phi ((+\frac 12)^{(\widehat{{\bf b}})};(+\frac
12)^{(\widehat{{\bf c}})})\right| ^2+\left| \phi ((+\frac 12)^{(\widehat{%
{\bf b}})};(-\frac 12)^{(\widehat{{\bf c}})})\right| ^2]  \label{th36}
\end{equation}

\begin{eqnarray}
(\sigma ^2)_{12} &=&3[\phi ^{*}((+\frac 12)^{(\widehat{{\bf b}})};(+\frac
12)^{(\widehat{{\bf c}})})\phi ((-\frac 12)^{(\widehat{{\bf b}})};(+\frac
12)^{(\widehat{{\bf c}})})  \nonumber \\
&&+\phi ^{*}((+\frac 12)^{(\widehat{{\bf b}})};(-\frac 12)^{(\widehat{{\bf c}%
})})\phi ((-\frac 12)^{(\widehat{{\bf b}})};(-\frac 12)^{(\widehat{{\bf c}}%
)})],  \label{th37}
\end{eqnarray}

\begin{eqnarray}
(\sigma ^2)_{21} &=&3[\phi ^{*}((-\frac 12)^{(\widehat{{\bf b}})};(+\frac
12)^{(\widehat{{\bf c}})})\phi ((+\frac 12)^{(\widehat{{\bf b}})};(+\frac
12)^{(\widehat{{\bf c}})}c)  \nonumber \\
&&+\phi ^{*}((-\frac 12)^{(\widehat{{\bf b}})};(-\frac 12)^{(\widehat{{\bf c}%
})})\phi ((+\frac 12)^{(\widehat{{\bf b}})};(-\frac 12)^{(\widehat{{\bf c}}%
)})]  \label{th38}
\end{eqnarray}
and

\begin{equation}
(\sigma ^2)_{22}=3[\left| \phi ((-\frac 12)^{(\widehat{{\bf b}})};(+\frac
12)^{(\widehat{{\bf c}})})\right| ^2+\left| \phi ((-\frac 12)^{(\widehat{%
{\bf b}})};(-\frac 12)^{(\widehat{{\bf c}})})\right| ^2]  \label{th39}
\end{equation}

Using the Land\'e expansion Eq. (\ref{fo40a}) and the Hermiticity condition
Eq. (\ref{fo41a}), we see that 
\begin{eqnarray}
(\sigma ^2)_{11} &=&3[\phi ((+\frac 12)^{(\widehat{{\bf b}})},(+\frac 12)^{(%
\widehat{{\bf c}})})\phi ((+\frac 12)^{(\widehat{{\bf c}})};(+\frac 12)^{(%
\widehat{{\bf b}})})  \nonumber \\
&&\ +\phi ((+\frac 12)^{(\widehat{{\bf b}})},(-\frac 12)^{(\widehat{{\bf c}}%
)})\phi ((-\frac 12)^{(\widehat{{\bf c}})};(+\frac 12)^{(\widehat{{\bf b}}%
)})]  \nonumber \\
\ &=&3\zeta ((+\frac 12)^{(\widehat{{\bf b}})};(+\frac 12)^{(\widehat{{\bf b}%
})})=3,  \label{fo40}
\end{eqnarray}

\begin{eqnarray}
(\sigma ^2)_{12} &=&3[\phi ((-\frac 12)^{(\widehat{{\bf b}})};(+\frac 12)^{(%
\widehat{{\bf c}})})\phi ((+\frac 12)^{(\widehat{{\bf c}})};(+\frac 12)^{(%
\widehat{{\bf b}})})  \nonumber \\
&&+\phi ((-\frac 12)^{(\widehat{{\bf b}})};(-\frac 12)^{(\widehat{{\bf c}}%
)})\phi ((-\frac 12)^{(\widehat{{\bf c}})};(+\frac 12)^{(\widehat{{\bf b}}%
)})]  \nonumber \\
&=&3\zeta ((-\frac 12)^{(\widehat{{\bf b}})};(+\frac 12)^{(\widehat{{\bf b}}%
)})=0.  \label{fo41}
\end{eqnarray}
where $\zeta $ is the probability amplitude for measuring spin projections
twice in a row along the same quantization axis. Similarly, we find that 
\begin{equation}
(\sigma ^2)_{21}=0\text{ and }(\sigma ^2)_{22}=3.  \label{fo42}
\end{equation}
Hence,

\begin{equation}
\lbrack \sigma ^2]=3\left( 
\begin{array}{cc}
1 & 0 \\ 
0 & 1
\end{array}
\right) ,  \label{fo43}
\end{equation}
a result we obtained earlier by squaring the components of the spin and
adding them. Since this operator is the unit matrix multiplied by a
constant, the eigenvectors of $[\sigma _{\widehat{{\bf c}}}]$ are also
eigenvectors of the operator.

\section{Discussion and Conclusion}

We have presented the eigenvectors of the generalized operators of the $x$
and $y$ components of spin. More importantly, we have shown how the
components $[\sigma _x]$ and $[\sigma _y]$ can be obtained from the operator 
$[\sigma _{\widehat{{\bf c}}}]$ by transformation of the angles. We have
obtained $[\sigma _x]$ and its eigenvectors by the transformation $\theta
^{\prime }\rightarrow \theta ^{\prime }-\pi /2$ applied to $[\sigma _{%
\widehat{{\bf c}}}]$ and its eigenvectors . Similarly, we have obtained $%
[\sigma _y]$ and its eigenvectors by the transformations $\theta ^{\prime
}=\pi /2$ and $\varphi ^{\prime }\rightarrow \varphi ^{\prime }-\pi /2.$
This procedure for obtaining the $x$ and $y$ components of the spin operator
from the $z$ component should be of general validity and should therefore
save labour when the generalized operators and their eigenvectors for higher
values of spin are sought.

We have also proved that the operator for the square of the spin is the unit
matrix multiplied by the eigenvalue of the square of the spin. Every
two-dimensional vector is an eigenvector of this operator, so that the
requirement that the eigenvectors of $[\sigma _{\widehat{{\bf c}}}]$ also be
the eigenvectors of $[\sigma ^2]$ is automatically satisfied.

We have now completed the task of deriving generalized operators and vectors
for spin $1/2$ using the Land\'e interpretation of quantum mechanics. The
next task is to extend this approach to spin $1,$ both as a way of verifying
the validity of the new approach, and to obtain the generalized operators
and their eigenvectors for spin $1$.

\section{References}

1. Mweene H. V., ''Derivation of Spin Vectors and Operators From First
Principles'', not yet published but submitted to {\it Foundations of Physics.%
}, quant-ph/9905012.

2. Land\'e A., ''New Foundations of Quantum Mechanics'', Cambridge
University Press, 1965.

3. Land\'e A., ''From Dualism To Unity in Quantum Physics'', Cambridge
University Press, 1960.

4. Land\'e A., ''Foundations of Quantum Theory,'' Yale University Press,
1955.

5. Land\'e A., ''Quantum Mechanics in a New Key,'' Exposition Press, 1973.

\end{document}